\documentclass[prd,aps,twocolumn,nofootinbib,preprintnumbers]{revtex4}
\usepackage{amsmath,amssymb,epsf}
\usepackage{graphicx}

\begin{document}

\title { \Large  \bf  Isolated Minkowski vacua, and stability analysis \\for an extended brane in the rugby ball}
\author{Burak Himmeto\={g}lu and Marco Peloso}
\affiliation{School of Physics and Astronomy \\
University of Minnesota, Minneapolis, MN 55455, USA}
\date{\today}

\begin{abstract}
We study a recently proposed model, where a codimension one brane
is wrapped around the axis of symmetry of an internal two dimensional
space compactified by a flux. This construction is free from the problems
which plague delta-like, codimension two branes, where only tension can be present.
In contrast, arbitrary fields can be localized on this extended
brane, and  their gravitational interaction is standard $4$d
gravity at large distance.  In the first part of this note, we study the de Sitter (dS)
vacua of the model. The landscape of these vacua is characterized by
discrete points labeled by two integer numbers, related to the flux responsible for
the compactification and to the current of a brane field. A Minkowski external space emerges only
for a special ratio between these two integers, and it is therefore (topologically) isolated from 
the nearby dS solutions. In the second part, we show that the Minkowski vacua are stable
under the most generic axially-symmetric perturbations (we argue that this is sufficient to ensure the overall stability). 
\end{abstract}

\preprint{{\tt hep-th/0612140}}
\preprint{UMN-TH-2531/06}

\maketitle

\section{Introduction}

Models with two extra dimensions have been the object of 
many studies. Among them, there are the first examples of
supersymmetric compactifications to four dimensional
Minkowski space~\cite{Salam}, where a stable spherical internal space is achieved
through the flux of a gauge field and a cosmological constant.
 More recently, six dimensional
models have been reconsidered in the brane-world scenario
\cite{add}, since two is the minimum number of flat extra
dimensions to have a fundamental scale of order TeV and a
compactification of (sub)millimeter range. 
The background solution of~\cite{Salam} can be easily extended to a $6$d braneworld. If one imposes that the internal and the external coordinates are factorized, then the presence of a brane modifies the internal geometry by creating a deficit angle~\cite{self1}, in the same way as a point mass modifies a two dimensional space~\cite{djth}. In this way, the spherical internal space of~\cite{Salam} is modified
to a (so called) rugby ball, with conical singularities at the two poles generated by the codimension
two brane and its $Z_2$ image. It is manifest from the solution that the only quantity
related to the brane tension is the deficit angle in the bulk, so that
one may hope that this construction can provide a self tuning
mechanism responsible for the vanishing of the $4$d cosmological constant.
Unfortunately, this is not the case for a number of reasons~\cite{self2}.
Nonetheless, the construction of \cite{self1} remains of great interest, since
it is an extremely simple but complete solution of unwarped extra
dimensions, which can be employed in a number of applications of
physical relevance \cite{fieldtheory6}.

There is, however, a further problem that these applications face,
related to the localization of matter and gauge fields on the
codimension two branes. In field theoretical studies, matter and
gauge fields on the brane are usually treated as test fields, which
do not contribute to the background geometry (namely, they are
neglected in the Einstein equations of the system). However, a
more complete approach, with gravity also taken into
account, shows that only tension can be present on the brane,
while fields with a different equation of state necessary lead to
worse than conical singularities at the brane location
\cite{CBK}. Contrary to what happens for conical singularities
(where the scalar curvature is a delta function supported at the tip
of the cone) such singularities are not integrable, so that the 
codimension two brane cannot be consistently
treated even in a distributional sense. This appears as a specific example 
of the problem of defining sources of codimension two and higher 
in general relativity \cite{gt} (see also~\cite{nemanja} for a recent discussion
focused on exact codimension $2$ solutions).

A way out of this problem was proposed
in~\cite{CL}, through the addition of
Gauss-Bonnet terms in the bulk, and in~\cite{Kaloper}, where the
codimension two brane emerges at the intersection between two
codimension one branes. A possibly more conservative approach
is to replace the delta-like strict codimension two brane with an extended
codimension one defect~\cite{PST} (extended defects on a
codimension two bulk were also discussed in~\cite{ext}). In this
model, a region close to the singularity is replaced by a
spherical cap, and a codimension one defect is placed at the
junction between the two spaces (the same is done for both
singularities, so to preserve the $Z_2$ symmetry of the system;
moreover, the defect wraps around the main axis of symmetry of the
system; the overall configuration can be seen in fig. 1 of~\cite{PST}). 
Fields with arbitrary equation of state can be localized
on the defect, and their zero modes (that is, the axially-symmetric
ones) can be identified with the $4$d fields of the
observable sector. It was shown in \cite{PST} that the gravitational interaction
between brane fields is described by Einstein $4$d gravity at large
distance. Therefore, this construction appears as a
phenomenologically viable and complete theory of two extra
dimensions, without the need of invoking higher order gravity
terms, or a more complicated system of branes.

A natural extension of this construction is the study of more general background solutions. For instance, ref.~\cite{PPZ} embedded the codimension one brane in a Minkowski compactification characterized by a warped internal space. Alternatively, one can study cases of cosmological relevance, where the brane is embedded in a time dependent background. We perform the first step in this direction by studying and classifying the de
Sitter (dS) vacua of this model. More precisely, we look for
solutions characterized by a dS external geometry, and a static
internal space. More general solutions can presumably
be obtained analytically for small deviations from the dS or Minkowski ones; in 
general, the values of the bulk cosmological constants and of the brane tension determine both
the dS expansion rate, and the bulk parameters, such as the compactification 
radius and the deficit angle; if more general sources are present, we 
expect a nearly static internal space, and an approximate Friedmann-Robertson-Walker cosmology, as long as their energy densities are smaller than the brane and bulk cosmological constants (this is typical of extra dimensional models with a stabilized internal space). At  higher energies, we expect a strong evolution of the entire space, which can be presumably studied only through numerical computations~\cite{branecode}.

We find two main results: first, the dS rate
must be smaller than a given value, related to the inverse of the
compactification radius. This is in agreement with the findings
of~\cite{dsbra}, where it was observed that any
compactification mechanism can stabilize the internal space only
up to some given expansion rate (by causality reasons, one cannot
expect compactification when the horizon size becomes
parametrically smaller than the size of the internal space). A
second, more surprising, result, is that the Minkowski solutions, obtained
for a special relation between the bulk cosmological constants and the brane tension,
are separated from the dS ones by topology.
Indeed, the possible vacua of the model are labeled by two integer numbers.
The first integer $N$ is related to the winding number of bulk fields charged under the gauge
symmetry responsible for the compactification of the space (if present, these fields impose a quantization condition analogous to the one taking place for the Dirac monopole). This can also be seen as quantization of the flux of the gauge field in the compact space. The second integer $n$ is related to the current of a brane field; this current is necessary for matching the discontinuity of the gauge field across the two sides of the brane, and it controls the position of the brane in the internal space. Ref.~\cite{PST} studied only the Minkowski solutions for the system, showing that they occur for 
$N / n = - 2 \,$. We find that a nonvanishing dS rate is instead achieved for greater ratios. In the Conclusions, we comment on the implications that this can have for the cosmological constant problem.

Clearly, for the construction of~\cite{PST} to be of any interest, one should show that it is stable.
Only the zero modes of the perturbations of this geometry
were studied in \cite{PST}. In the second part of this work, we study the massive
modes, to ensure that the system has no tachyonic instability (we do so only for the Minkowski vacua; from what we already mentioned, and from the stability study for the spherical compactification without branes~\cite{jo}, we expect that the system becomes unstable at high $H$). More precisely, we concentrate only on axially symmetric perturbations (around the axis of symmetry of the background). 
The reason is mostly technical, since, as we will
see, already this  system of modes is quite involved. However, we
have a second (although non rigorous) justification for this
restricted study. The construction is obtained by cutting in an
axially symmetric way the rugby ball and the spherical
compactifications, and by joining them across the brane. Before
cutting them, both these configurations are stable~\cite{Salam}, so we expect
that an instability - if any - will be related to their interface.
For instance, the rugby ball may prevail over the spherical cap,
so that the brane in between them would shrink towards the pole. 
Alternatively, the spherical cap may be favored, and the string would
then extend towards the equator. Such instability would show up as a tachyonic
axially symmetric mode; recently, the stability of
$6D$ chiral gauged supergravity, including the unwarped ``rugby--ball'' solution,
was studied in~\cite{burgess}. Also that analysis is restricted to
axially symmetric perturbations; it is argued that the study of more general modes is
unnecessary, since any angular dependence would contribute
positively to the corresponding Kaluza-Klein mass.

We decompose the perturbations into scalar, vector, and tensor modes
(where the names refer to how they transform with respect to
transformations along the noncompact coordinates). These three
sectors are decoupled at the linearized level, and they can be
studied separately. The equations for the zero modes studied
in~\cite{PST} could be solved analytically. Unfortunately, for
massive modes, only the tensor mode can be obtained analytically,
while the bulk equations for the vector and scalar modes have to
be solved numerically.~\footnote{We are aware of two studies which are close to the present one,
where the equations in the bulk could be decoupled, and then studied analytically. 
For a spherical bulk, the modes can be decomposed on spherical harmonics, and then decoupled due to orthogonality in the bulk~\cite{jo}. This cannot be done for the ``composite'' bulk solution that we are investigating. Second, in the study of~\cite{burgess} the modes could be decoupled by using the equation for the dilaton present in the supergravity action. This mode, and the corresponding equation, is absent in our case (for other studies of linearized gravity in $6d$ contexts, see~\cite{flutt}).} We do so with a shooting method. No
tachyonic solution emerged from the (rather extensive) numerical computation, so that
we can conclude 

The paper is organized as follows. In Section 2 we review the
model introduced in~\cite{PST}. Section 3 is devoted to the study of the dS backgrounds. 
In Section 4 we show that the system is stable under the the most general set of axially symmetric perturbations. The concluding Section 5 contains some remarks on the phenomenological aspects of the model,  with a particular focus on the possible implications of the separation between the Minkowski and dS solutions.

\section{Minkowski compactification}

We summarize the construction of~\cite{PST}. The
action of the model is
\begin{eqnarray}
&& S=S_{o}+S_{i}+S_{b} \\
&& S_{o,i}=\int d^{6}x\sqrt{-g_{6}}\left[\frac{M^4}{2}R-
\Lambda_{o,i} -\frac{1}{4} F^2 \right]
\nonumber \\
&& S_{b}=-\int
d^{5}x\sqrt{-\gamma}\left[\lambda_{s}+\frac{v^2}{2}\left(\partial \sigma-eA \right)^2
\right] \nonumber
\end{eqnarray}
where $S_b$ is the action of the brane, while $S_{o,i}$ is the action in two bulk regions separated by the brane. The line element in the two regions is
\begin{eqnarray}
&&
ds_{6}^{2}=\eta_{\mu\nu}dx^{\mu}dx^{\nu}+R^2d\theta^{2}+R^2\beta^{2}\cos^{2}\theta
d\phi^{2}
\quad \mathrm{"out"} \nonumber \\
&&
ds_{6}^{2}=\eta_{\mu\nu}dx^{\mu}dx^{\nu}+R^2\beta^{2}d\theta^{2}+R^2\beta^{2}\cos^{2}\theta
d\phi^{2} \quad \mathrm{"in"} \nonumber\\ \label{line}
\end{eqnarray}
where here and in the following $x$ denotes the noncompact directions (notice the choice of the Minkowski metric; in the next Section we will instead consider a dS external space). The brane is at the background position ${\bar \theta} \,$. The region at $0 < \theta < {\bar \theta}$, denoted as the ``out'' bulk, is a portion of the so call rugby ball compactification, characterized by the deficit angle $1 - \beta \,$. The region ${\bar \theta} < \theta < \pi /2$, denoted as ``in'' bulk, is a spherical cap, with the pole at $\theta = \pi /2$. A $Z_2$ symmetry extends this geometry to the region $- \pi/2 < \theta < 0 \,$. In addition, the background is axially symmetric (around the axis connecting the poles at $\theta = \pm \pi/2$).

The internal compactification is achieved through the gauge field configuration
\begin{equation}
F_{\theta\phi}=\partial_{\theta}A_{\phi}=M^2 R \beta\cos\theta
\label{F}
\end{equation}
and the cosmological constants
\begin{equation}
\sqrt{2\Lambda_{i}}=\frac{M^2}{R\beta} \;\;,\;\;
\sqrt{2\Lambda_{0}}=\frac{M^2}{R} \label{Lambda}
\end{equation}

If we suppress the external dimensions, the brane can be viewed as a string wrapped
around this axis of symmetry. The brane field $\sigma$ acts as a Goldstone boson ($v$ is most easily interpreted as the vacuum expectation value of a field which breaks the $U(1)$ symmetry on the brane). It generates a current which is necessary to provide the discontinuity of the magnetic field between the two bulk regions. From its own equation of motion, $\sigma = n \, \phi$, where (due to the periodicity of the $\phi$ coordinate) $n$ is an integer. The brane position is then found to be~\cite{PST}
\begin{equation}
{\bar \theta} = \arctan \, \frac{1-\beta}{R \beta q^2} \;\;\;,\;\;\; q \equiv e \, v
\end{equation}
while the deficit angle in the out bulk is
\begin{equation}
1 - \beta = \frac{T}{2 M^4 \pi \sin {\bar \theta}}
\end{equation}
where $T$ is the four dimensional (i.e., after an integration along $\phi$) energy density of the brane. This result, in the limit ${\bar \theta} \rightarrow \pi /2$ (when the brane shrinks to the north pole), reproduces the known relation between the tension of a codimension two brane, and the deficit angle generated by it.

To conclude the description of the background, we note that, if some field, with charge $e$ under the $U(1)$ symmetry is present, the deficit angle must satisfy a quantization condition
\begin{equation}
\beta = \frac{N}{2 e M^2 R} \;\;\;,\;\;\; N \; {\rm integer}
\label{nh0}
\end{equation}
Such quantization is also known as flux quantization, since it can be recast in the form
\begin{equation}
\Phi_B \equiv \int d \theta \, d \phi \, F_{\theta \phi} = \frac{2 \pi}{e} \, N
\label{quantflux}
\end{equation}
where the integral is performed over the entire internal space.

From the Einstein equations, one then finds \cite{PST} that
this integer $N$ must be related to the winding $n$ of the brane field $\sigma$
by $N/n = - 2 \,$. As we show in the next Section, this relation is
actually due to the assumption of Minkowski noncompact space.
Different ratios between these two integers result in a dS
external geometry.

\section{de-Sitter Compactification}

We now generalize the Minkowski solution described above to the case of a dS noncompact
space, characterized by the expansion rate $H \,$ (namely, we replace $\eta_{\mu \nu}$ by the dS metric in the line elements~(\ref{line})). The bulk compactification is achieved for
\begin{eqnarray}
\Lambda_0 &=& \frac{M^4}{2 R^2} \left( 1 + 9 H^2 R^2 \right) \nonumber\\
F_{\theta\phi} &=& M^2R\beta\sqrt{1-3H^2R^2} \, \cos\theta 
\label{bds}
\end{eqnarray}
in the out bulk, and
\begin{eqnarray}
\Lambda_i &=& \frac{M^4}{2 R^2 \beta^2} \left( 1 + 9 \beta^2 H^2 R^2 \right) \nonumber\\
F_{\theta\phi} &=& M^2R\beta\sqrt{1-3H^2R^2\beta^2}\cos\theta 
\end{eqnarray}
in the in bulk. We observe that the expansion rate cannot exceed the value of $1/ \left( \sqrt{3} R \right)$.
This is not surprising in the light of the findings of~\cite{dsbra}, where it was shown that, for any dS compactification, the expansion of the external coordinates has the generic effect of destabilizing the internal space.

We still look for an axially symmetric solutions, so that $A_\theta = 0 \,$, while $A_\phi$ depends only on $\theta \,$. Moreover, $A_\phi =0$ at the poles (from regularity), and  $A_\phi$ is continuous across the two branes (so that the brane action is well defined). This determines $A_\phi$ in the bulk;  the solution cannot be provided on a unique chart. In presence of a bulk field charged under this $U(1)$ symmetry, a consistent solution is possible only when the quantization condition
\begin{eqnarray}
&& \!\!\!\!\!\!\!\!\!\!\!\!\!\!\!\!\!\!\!\!\!\!\!\! 2 e M^2 R \beta \left\{ \left( 1 - 3 H^2 R^2 \right)^{1/2} \sin
{\bar \theta} \right. \nonumber\\
&&\left. +  \left( 1 - 3 H^2 \beta^2 R^2 \right)^{1/2} \left(
1 - \sin {\bar \theta} \right) \right\} = N
\label{quant-H}
\end{eqnarray}
with $N$ integer, holds (this computation can be performed exactly as in the Minkowski case; see~\cite{PST} for details). Also in this case, the quantization condition can be recast in the form~(\ref{quantflux}).

Let us now discuss the brane equations. As for the Minkowski case, $\sigma = n \, \phi \,$, where $n$ is an integer. By construction, the transverse metric components ($g_{\mu \nu}$ and $g_{\phi \phi}$) are already continuous across the brane. We are then left with three nontrivial brane equations (two second Israel conditions, plus Ampere law, relating the discontinuity of $F_{\theta \phi}$ in the bulk to the current on the brane). With some algebra, they can be recast in the form
\begin{eqnarray}
&& \frac{1-\beta}{R\beta} = q^2 \tan{\bar\theta} \, {\cal F}^2
\label{eqn1}\\
&& 1+\frac{2n}{N} = \label{eqn2}\\
&& = \frac{\sin{\bar\theta}(\sqrt{1-3H^2R^2}- {\cal F})}{\sqrt{1-3H^2R^2\beta^2}(1-\sin{\bar\theta})+\sqrt{1-3H^2R^2}\sin{\bar\theta}} \nonumber\\
&& \frac{1-\beta}{R\beta}\tan{\bar\theta}=\frac{2\lambda_{s}}{M^4}
\label{eqn3}
\end{eqnarray}
where $q \equiv  e \, v \,$, and we have defined, for shortness,
\begin{equation}
{\cal F}  \equiv \frac{1 - \beta}{\left( 1 - 3 H^2
\beta^2 R^2 \right)^{1/2} - \left( 1 - 3 H^2 R^2 \right)^{1/2}
\beta}
\end{equation}

To solve the above system of equations, we eliminate ${\bar \theta}$ from (\ref{eqn1}) and (\ref{eqn3}), and we combine the resulting equation with the two bulk expressions for $\Lambda_0$ and $\Lambda_i$. 
We obtain
\begin{eqnarray}
1 - \beta^2 &=& \frac{16 \lambda q^2 \left( \Lambda_i - \Lambda_0 \right)}{3 \left( \Lambda_i - \Lambda_0 \right)^2 + 2 \lambda q^2 \left( 5 \Lambda_i - 3 \Lambda_0 \right)  + 3 \lambda^2 q^4} \nonumber\\
R^2 &=& \frac{16 \lambda q^2}{3 \left( \Lambda_i - \Lambda_0 \right)^2 - 2 \lambda q^2 \left( 3 \Lambda_i - 5 \Lambda_0 \right)  + 3 \lambda^2 q^4} \nonumber\\
H^2 &=& \frac{- \left( \Lambda_i - \Lambda_0 \right)^2 + 2 \lambda q^2 \left( \Lambda_i + \Lambda_0 \right) - \lambda^2 q^4}{48 \lambda q^2}
\label{dssol}
\end{eqnarray}
where we have rescaled
\begin{equation}
\frac{2 \Lambda_0}{M^4} \rightarrow \Lambda_0 \;\;,\;\;
\frac{2 \Lambda_i}{M^4} \rightarrow \Lambda_i \;\;,\;\;
\frac{2 \lambda_s}{M^4} \rightarrow \lambda 
\end{equation}

These solutions are valid only for $\lambda \leq \left( \Lambda_i - \Lambda_0 \right) / q^2 \,$ (this is because they are obtained by squaring some of the above equations). The maximal allowed value leads  to $H = 1 / \left( \sqrt{3} R \right) \,$, which, as we saw from eq.~(\ref{bds}), is the highest possible value that the system can have for the dS expansion rate. 

We see that the Minkowski compactification requires the tuning
$  
\sqrt{\lambda} = \left( \sqrt{\Lambda_i}-\sqrt{\Lambda_0} \right) /  
q
\,$.  For a small deviation
\begin{equation}
\sqrt{\lambda} = \frac{\sqrt{\Lambda_i}-\sqrt{\Lambda_0}}{q} + \delta
\label{delta}
\end{equation}
eqs.~(\ref{dssol}) give
\begin{eqnarray}
\beta^2 &=& \frac{\Lambda_0}{\Lambda_i} - \frac{3 \sqrt{\Lambda_0} \left( \sqrt{\Lambda_i} + \sqrt{\Lambda_0} \right) q}{2 \Lambda_i^{3/2}} \delta + {\rm O} \left( \delta^2 \right) \nonumber\\
R^2 &=& \frac{1}{\Lambda_0} + \frac{3 \sqrt{\Lambda_i} \, q}{2 \Lambda_0^{3/2} \left( \sqrt{\Lambda_i} - \sqrt{\Lambda_0} \right)} \delta + {\rm O} \left( \delta^2 \right) \nonumber\\
H^2 &=& \frac{\sqrt{\Lambda_0} \sqrt{\Lambda_i} \, q}{6 \left( \sqrt{\Lambda_i} - \sqrt{\Lambda_0} \right)} \delta - \frac{q^2 \left( \Lambda_i + \sqrt{\Lambda_i} \sqrt{\Lambda_0} + \Lambda_0 \right)}{12 \left( \sqrt{\Lambda_i} - \sqrt{\Lambda_0} \right)^2} \delta^2 \nonumber\\
&&+ {\rm O} \left( \delta^3 \right)
\label{exp}
\end{eqnarray}
From either of (\ref{eqn1}) and (\ref{eqn3}), the brane position then satisfies
\begin{eqnarray}
\tan {\bar \theta} &=& \frac{\sqrt{\Lambda_i} - \sqrt{\Lambda_0}}{q^2} + \frac{5}{4 \, q} \delta  + {\rm O} \left( \delta^2 \right)
\label{tat}
\end{eqnarray}

The possible values for the above parameters are constraint by the two integer values $N$ and $n \,$. 
Eqs.~(\ref{quant-H}) and~(\ref{eqn2}) give
\begin{eqnarray}
N &=& 2 \, e \, M^2 / \sqrt{\Lambda_i} + {\rm O} \left( H^2 \right)  \label{quant} \\
1 + \frac{2 n}{N} &=& - \frac{3 \left( \sqrt{\Lambda_i} - \sqrt{\Lambda_0} \right)^2}{2 \Lambda_0 \sqrt{\Lambda_i} \sqrt{\left( \sqrt{\Lambda_i} - \sqrt{\Lambda_0} \right)^2 + q^4}} H^2 + {\rm O} \left( H^4 \right) \nonumber
\end{eqnarray}
where the expansion in $\delta$ has been replaced by an expansion in $H^2$ through the last of~(\ref{exp}).

From the first quantization condition, we see that we cannot vary $\lambda$ alone without also varying the bulk cosmological constants. However, the Minkowski compactification ($\delta = 0$) can be still
deformed continuously into a dS one. However, once also the second condition is taken into account, the Minkowski solution appears to be detached from the dS ones. Indeed, the choice $2 n / N = - 1$
is only compatible with $H = 0 \,$, with the only exception of the trivial case of $\Lambda_i = \Lambda_0$ (in which case, $\beta =1$, and the brane is actually absent).~\footnote{This conclusion actually holds for arbitrary values of $H \,$, and not just at small $\delta$, as can be seen by studying eq.~(\ref{eqn2}) for $2 n / N = - 1$.} 

We can gain further insight by estimating the parameters entering in eq.~(\ref{quant}). Neglecting the 
subleading $H^2$ terms, and for non hierarchical values of the deficit angle (that is, $\beta$ and $1 - \beta$ of order one), both $\sqrt{\Lambda_i}$ and $\sqrt{\Lambda_0}$, as well as their difference, are of order $1/ R^2 \,$. In addition, $q^2 R$ does not exceed one (as can be seen in eq.~(\ref{tat}) - this value controls the ratio between the radius of the brane and that of the internal space). Therefore, we find
\begin{equation}
1 + \frac{2 n}{N} = {\rm O} \left( R^2 \, H^2 \right) 
\label{dsratio}
\end{equation}
From the same reasoning, we also see that the expansion at small $\delta$ is actually an expansion for $R \, H \ll 1 \,$. 

In the concluding Section we comment on the implication of these findings for the cosmological constant problem.

\section{Stability of the Minkowski compactification}

The goal of this Section is to obtain the massive perturbations of the system, to verify whether the background solution described in Section $2$ has tachyonic instabilities. For the reasons mentioned in the Introduction, we focus on axially symmetric perturbations. ~\footnote{Although involved, it is not hard to extend this analysis to general modes. Since the background is axially symmetric, the general dependence of the modes on the angular coordinate can only be of the form ${\rm exp}  \left( i \, n \, \phi \right) \,$, with $n$ integer.} The most general perturbations of the geometry of this type are
\begin{eqnarray}
&& ds_{6}^{2}=\left(1+2\Phi\right)dl^2+2A dl d\hat\phi+
\left(1+2C\right)\cos^{2}\theta d\hat\phi^2+ \nonumber \\
&& \qquad +2\left(T_{\mu}+\partial_{\mu}T\right)dl
dx^{\mu}+2\left(V_{\mu}+\partial_{\mu}V\right)d\hat\phi dx^{\mu}+
\nonumber \\
&& \qquad
\left\{\eta_{\mu\nu}\left(1+2\Psi\right)+2E_{,\mu\nu}+E_{(\mu,\nu)}+h_{\mu\nu}\right\}dx^{\mu}dx^{\nu}
\nonumber\\
\label{perturb}
\end{eqnarray}
where we have defined the ``dimensionful angular coordinates''
\begin{equation}
dl \equiv \left\{ \begin{array}{l}
R\beta d\theta \quad \mathrm{"in"} \\
R d\theta \quad \mathrm{"out"}
\label{defns}
\end{array} \right., \qquad d\hat\phi \equiv R\beta d\phi \;\;,\;\;
\end{equation}
and $E_{(\mu,\nu)}\equiv\partial_{\nu}E_{\mu}+\partial_{\mu}E_{\nu}$. The vector modes $E_{\mu}$, $T_{\mu}$, $V_{\mu}$ are transverse, and the
tensor mode $h_{\mu\nu}$ is transverse and traceless. The remaining modes are scalar (the
denomination refers to how these modes transform under $4$d coordinate transformations).
Simultaneously, one needs to consider the perturbations of the gauge field,
\begin{equation}
\delta A_{\hat\phi}=a_{\hat\phi}, \qquad \delta A_{l}=a_{l},
\qquad \delta A_{\mu}=\partial_{\mu}a+\hat{a}_{\mu} \label{pertA}
\end{equation}
where $\hat{a}_{\mu}$ is a transverse vector mode. Following the discussion of \cite{PST}, we fix part of gauge freedom by setting $E_{\mu}=T=V=a_l-a'=0$ (here and in the following, prime denotes derivative with respect to the rescaled coordinate $l \,$). We further impose
\begin{equation}
\theta_{\rm brane} = {\bar \theta} \;\;\;,\;\;\; \Phi \left( {\bar \theta} \right) = 0
\end{equation}
that is, we require that the brane lies at the unperturbed background position, and that $g_{ll} = 1$ there (this choice includes the Gaussian normal coordinate choice at the brane location, which is the most convenient one to interpret the gravitational effects measured by brane observers). These choices do not fix the gauge completely (see~\cite{PST} for details); however, we can still have a general (and unambiguous) study if we perform our computation in terms of the combination of modes which are invariant under the residual gauge freedom.

The tensor mode $h_{\mu \nu}$, and the vector modes $T_\mu ,\, {\hat a}_{\mu} ,\,$ and $V_\mu \,$, are already invariant. For the scalar sector, the invariant combinations are instead
\begin{eqnarray}
&& \hat\Phi=\Phi+E''  \nonumber \\
&& \hat{C}=C-\theta'\tan\theta E' \nonumber \\
&& \hat{a}_{\hat\phi}=a_{\hat\phi}+M^2\theta'\cos\theta E'  \nonumber \\
&& \Psi
\label{hatsca}
\end{eqnarray}

Tensor, vector and scalar modes are decoupled at the linearized level, so we can study
the three sectors separately. We do so in the next three Subsections. The relevant equations were obtained in~\cite{PST}, where the zero modes of the system were then studied. The derivation of these equations is not repeated here.

\subsection{Tensor Modes}

The axially symmetric tensor perturbation can be decomposed as
\begin{equation}
h_{\mu \nu} \left( x ,\, \theta \right) = \sum_n h_n \left( \theta \right) \, C_{\mu \nu ,\,n} \left( x \right)
\label{deco}
\end{equation}
where $C_{\mu \nu ,\,n}$ are $4$d Kaluza Klein (KK) tensor modes, and $h_n$ their wavefunctions in the bulk. Our goal is to find the allowed perturbations, and their $4$d masses $m_n \,$. The bulk equation
\begin{equation}\label{tensor1}
\partial^2 h_{\mu\nu}+h_{\mu\nu}''-\theta' \tan\theta
h_{\mu\nu}'=0
\end{equation}
(where $\partial^2$ denotes the d'Alambertian operator in $4$d) can be separated in
\begin{eqnarray}
&&\frac{d^2 h_n}{d \theta^2} -\tan\theta \frac{d h_n}{d \theta}+\mu_n^2 h_n=0 \nonumber\\
&& \partial^2 C_{\mu \nu ,\, n} = m_n^2 C_{\mu \nu ,\, n}
\end{eqnarray}
where we have moved back to the $\theta$ coordinate, and where the parameter $\mu_n$ is different in the two bulk regions:
\begin{eqnarray}
\mu_n^2 \equiv \left\{
\begin{array}{l}
m_n^2 R^2\beta^2 \quad \mathrm{"in"} \\
m_n^2 R^2 \quad \mathrm{"out"}
\end{array} \right. 
\label{eqh}
\end{eqnarray}
From now on, we suppress the index $n$ for brevity, understanding that we are studying one KK mode at a time.

The bulk equations must be supplemented by a set of boundary and
parity conditions. First, we require regularity at the two poles,
imposing that the first derivative of $h$ vanishes there (we can
impose this condition either on the derivative
with respect to $\theta$ or $l$, since the two variables are
simply related by a constant rescaling). Second, parity
considerations impose that the modes are even across the equator,
$h \left( - \theta \right) = h \left( \theta \right) \,$. Finally,
we must satisfy the junction conditions across the brane,
\begin{eqnarray}
&& h_{in}(\bar\theta) = h_{out}(\bar\theta) \nonumber\\
&& h_{in}'(\bar\theta)=h_{out}'(\bar\theta)  \;\;\; {\rm or} \;\;\;
\frac{\partial h_{in}}{\partial \theta} (\bar\theta)=\beta \frac{\partial h_{out}}{\partial \theta}
(\bar\theta)
\label{tensBC}
\end{eqnarray}

The bulk equations are solved by the Legendre functions $P_{\nu}(x)$ and
 $Q_{\nu}(x)$, where $x=\sin\theta$ and $\nu=-1/2+\sqrt{1/4+\mu^2}$
(we denote by $\nu_{i}$ and $\nu_{o}$ the values of this parameter in the in and out bulk,
respectively). The bulk solution which is regular at the poles, even across the equator,
and satisfies the first of~(\ref{tensBC}) is
\begin{eqnarray}
&& h_{in}=P_{\nu_{i}}\left(|x|\right) \nonumber\\ 
&& h_{out}=A \, \left[\cos\left(\frac{\pi\nu_{o}}{2}\right)P_{\nu_{o}}(x) - \frac{2}{\pi}\sin\left(\frac{\pi\nu_{o}}{2}\right)Q_{\nu_{o}}(x)\right]
\nonumber\\
\end{eqnarray}

where
\begin{equation}
A=\frac{P_{\nu_{i}}(|\bar
x|)}{\cos\left(\frac{\pi\nu_{o}}{2}\right)P_{\nu_{o}}(\bar
x)-\frac{2}{\pi}\sin\left(\frac{\pi\nu_{o}}{2}\right)Q_{\nu_{o}}(\bar
x)} \;\;\;,\;\;\; \bar x=\sin\bar\theta
\label{Hconst}
\end{equation}

The only undetermined parameter is the mass square of the mode,
which enters in the two parameters $\nu_{i,o}$.~\footnote{There
should also be an overall normalization, which cannot be
determined by the linearized system; we have fixed it by setting
$h=1$ at the two poles.} It can be found by imposing the only
remaining condition to be satisfied, namely the second
of~(\ref{tensBC}). Specifically, for any fixed values of $\beta$
and ${\bar \theta}$, we (numerically) look for the roots of
\begin{equation}
f \left( m^2 \right) = \frac{\partial}{\partial x} \left[ h_{\rm in} - \beta h_{\rm out} \right]_{\big\vert_{\bar x}}
\end{equation}

\begin{figure}
\centering
\includegraphics[width=3in]{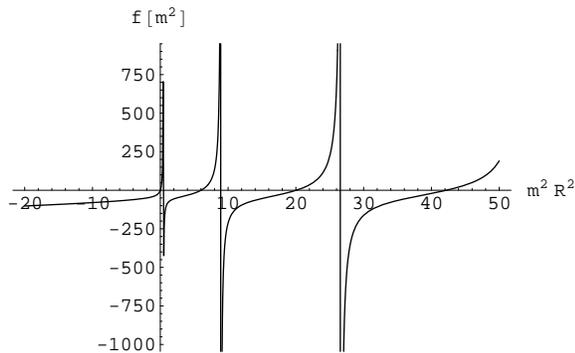}
\caption{f($\nu(m^2)$)
for the specific choice $\beta = 0.9$ and ${\bar \theta} = 85^0$.
} 
\label{fig1-1}
\end{figure}

\begin{figure}
\includegraphics[width=3in]{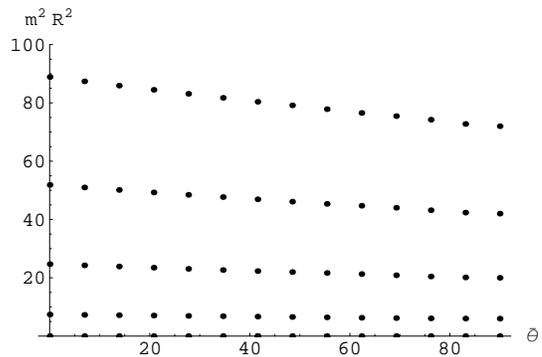}
\caption{Smallest masses for the tensor modes,
for the specific choice
$\beta = 0.9$ and for different values of the  brane
position}\label{fig1-2}
\end{figure}

As an example, Fig .(\ref{fig1-1}) shows the behavior of $f
\left(m^2 \right)$ for the specific choices of $\beta=0.9$ and
$\bar\theta=85^0$. As can be seen in the figure, there are no
tachyonic modes in the spectrum (this is the case also for more
negative values of $m^2$ than those plotted). We verified that no
tachyonic modes appear also for several other values of the brane
position (those reported in figure (\ref{fig1-2})) and the deficit
angle (We varied $\beta = 0.2 ,\, ... ,\, 0.9$ in steps of $0.1$ ). We also
notice the presence of the zero mode already studied
in~\cite{PST}.

Fig.(\ref{fig1-2}) shows instead the mass spectrum for
different brane positions, and for the specific choice
$\beta=0.9$. The points at ${\bar \theta} = 90^0$ have been
obtained for a codimension two brane located at the pole.
We see that
 the mass spectrum for the
codimension-1 model converges continuously to the one of
codimension-2 (we will see that this is not the case in the scalar sector).
This can be proven analytically, from the study of the equations which
determine the allowed modes. In the limit of ${\bar \theta} \rightarrow \pi / 2$,
the in part of the bulk shrinks to zero, so that all the bulk geometry is described by
the rugby ball, as it is the case for the codimension two case; so, the bulk equations
converge to that of codimension two. Moreover, the boundary/parity conditions that
we have discussed above reduce to
\begin{equation}
\frac{d h}{d \theta} _{\big\vert_{\pi / 2}} = 0 \;\;\;,\;\;\; h \left( - \theta \right) =
h \left( \theta \right) \;\;\;,\;\;\; {\rm as} \;\; {\bar \theta} \rightarrow \pi /2
\end{equation}
which coincide with those of the codimension two case.

In the codimension two case, the spectrum does not depend on $\beta$ (this is
strictly true for axially symmetric perturbations); this also emerges from our numerical results (not shown here): the dependence of the spectra on the deficit angle is very weak for any finite ${\bar \theta} \,$, and it disappears as the codimension one brane is shrunk to the pole.

\subsection{Vector Modes}

The linearized Einstein and Maxwell equations
for the vector modes have been derived in~\cite{PST}. The one for $T_\mu$ simply reads
\begin{equation}
\partial^2 T_\mu = 0
\end{equation}
which immediately indicates that this perturbation has not
massive modes. We decompose the two remaining modes as
we did for the tensor case,
\begin{equation}
{\hat a}_\mu = \sum_n {\hat \alpha}_{\mu \,, n} \left( x \right) \, a_n \left( \theta \right) \;\;,\;\;
{\hat V}_\mu = \sum_n w_{\mu \,, n} \left( x \right) \, W_n \left( \theta \right) / M^2
\end{equation}
 Omitting the index $n$ for brevity reasons, the bulk wavefunctions satisfy the following
system of equations
\begin{eqnarray}
&& \frac{d^2 a}{d \theta^2} - \tan \theta \, \frac{d a}{d \theta} + \mu^2 \, a - \frac{1}{\cos \theta} \, \frac{d W}{d \theta} = 0  \nonumber\\
&& \frac{d^2 W}{d \theta^2} + \tan \theta \, \frac{d W}{d \theta} + \mu^2 \, W + 2 \cos \theta \, \frac{d a}{d \theta} = 0
\label{bulkvec}
\end{eqnarray}
where $\mu$ is defined as in eq.~(\ref{eqh}).

Due to the parity choice of the background, the mode $a$ must be odd
across the equator, while the mode $W$ must be even,
\begin{equation}
a \left( - \theta \right) = - a \left( \theta \right) \;\;\;,\;\;\;
W \left( - \theta \right) = W \left( \theta \right)
\label{vectorBC2}
\end{equation}
In addition, there are
regularity conditions at the poles,
\begin{equation}
\frac{d a}{d \theta}_{| \pm \pi/2} = \frac{d W}{d \theta}_{| \pm \pi/2}  = 0
\label{vectorBC2-2}
\end{equation}
and junction conditions across the brane.
\begin{eqnarray}
&& a_{in}(\bar\theta)=a_{out}(\bar\theta) \;\;\;,\;\;\;
W_{in}(\bar\theta)=W_{in}(\bar\theta) \nonumber \\
&& \left(\frac{d a}{d \theta}-\frac{a}{\tan\bar\theta}\right)_{|\bar\theta \;, {\rm in}}=\beta\left( \frac{d a}{d \theta}-\frac{a}{\tan\bar\theta}\right)_{|\bar\theta \;, {\rm out}} \,,
\nonumber \\
&& \left( \frac{d W}{d \theta}+2a\cos\bar\theta\right)_{|\bar\theta \;, {\rm in}}=\beta\left( \frac{d W}{d \theta}+2a\cos\bar\theta\right)_{|\bar\theta \;, {\rm out}}  
\label{junctionsvec}
\end{eqnarray}

For massless modes, the bulk equations~(\ref{bulkvec}) form a
system of two coupled first order differential equations in terms
of $d a/d \theta$ and $d W/ d\theta$. This system can be solved
analytically. However, for nonvanishing mass, these equations must
be integrated numerically.

Therefore, to find the spectrum of vector modes, we resort to a
shooting method, which is appropriate for boundary value problems.
Each mode is specified
by its mass, and by a series of parameters which determine the
initial conditions at one of the poles. For definiteness, we start
from the south pole. As we discuss below, we actually need only
one such parameter, which we denote by ${\cal C}$. We start from
some guessed values for $m^2$ and ${\cal C}$, and we then solve
the bulk equations~(\ref{bulkvec}) (when we cross a brane, we
impose the conditions~(\ref{junctionsvec})). If the resulting
solution turns out to be regular, and to have the correct parity
assignment across the equator, then we have managed to identify
one physical mode of the system.

In practice, the bulk solutions that we obtain numerically
never satisfy these properties,
signaling that the initial guess for the parameters $m^2$ and
${\cal C}$ was wrong. We can define the two ``distances'',
\begin{equation}
d_{1} \equiv \left(\frac{a(-\bar\theta)+a(\bar\theta)}{a(-\bar\theta)-a(\bar\theta)}\right)
\qquad
d_{2} \equiv \left(\frac{W(-\bar\theta)-W(\bar\theta)}{W(-\bar\theta)+W(\bar\theta)}\right)
\label{numeric-dist}
\end{equation}
 which indicate how far the solution is from being $Z_2$ symmetric.
We then proceed in two steps: (i) we densely scan the parameter
space $\left\{ m^2 ,\, {\cal C} \right\}$ within some given range;
the values leading to the smallest distances are regarded as our
best guesses; (ii) we use a Newton's method to find the zeros of
these distances, starting from the best guesses. Provided the
initial conditions are dense enough, Newton's method converges to
all the physical solutions of the system, having values of $\left\{ m^2 ,\, {\cal C} \right\}$
not too far from the probed range of values.~\footnote{More 
accurately, the vanishing of $d_1$ and $d_2$
is a necessary, but not sufficient, condition for a mode to be a
physical perturbations of the system; indeed, in some cases those conditions were
(accidentally) satisfied, although the modes did not have the correct parity all
throughout the bulk. The easiest way to solve this problem
(in an automated way) is to check the parity condition also at
other bulk positions. Specifically, we discharged all those
solutions which satisfied $d_1 = d_2 = 0$, but which had the wrong
parity at the equator (physical solutions satisfy $W = d a / d
\theta = 0$ at $\theta = 0$). In all the cases we attempted, this
was enough to eliminate all the spurious solutions (we always
verified, by direct inspection, that all the modes which passed
this second check had the correct parity all throughout the bulk).}
Indeed, the two-dimensional nature of the initial parameter space,
and the fact that the bulk geometry is regular everywhere, make
the numerical problem a relatively simple one. It is easy to
verify (for instance, by increasing the density of the initial
scan) that all the solutions are reached with this method.

The main numerical difficulty occurs at the south pole, where the
coordinate system used is singular. To overcome this, we actually
solve (by Taylor expansion) the bulk equations analytically in a
neighborhood of the south pole. As for the tensor sector, there is
one overall normalization which cannot be determined by these
linearized equations. We fix this by imposing $a \left( - \pi/2
\right) = 1$ . 
~\footnote{This choice does not include the possibility of $a = 0$ at the pole.
In this case, the numerical computation only leads to the zero mode characterized by constant $a$ and $W$ in the entire bulk.} We then find
\begin{eqnarray}
a \left( - \pi/2 + \epsilon \right) &=& 1 + {\cal C} \, \epsilon^2 + {\rm O} \left( \epsilon^4 \right) \nonumber\\
W \left( - \pi/2 + \epsilon \right) &=& \left( 2 \, {\cal C} + \frac{\mu^2}{2} \right) \epsilon^2 + {\rm O} \left( \epsilon^4 \right)
\label{anpolvec}
\end{eqnarray}
The solutions of a system of second order equations, are usually specified by the values of the functions and their first derivatives at a given point. In the present case, due to the coordinate singularity at the poles, we also need to specify one of the second derivatives (we also note that the linear terms in the expansions vanish, due to the regularity conditions~(\ref{vectorBC2-2})). We started our numerical evolutions with $\epsilon = 10^{-6} \,$, with the values of the wave functions and their derivatives obtained from~(\ref{anpolvec}).

We performed the analysis for several values of $\beta$ (namely
$\beta=0.3, 0.6, 0.9$) and ${\bar \theta}$ ($\bar\theta=60, 65,
70, 75, 80, 85$). The initial region scanned was at $-100\leq
m^2\leq 100$ and $-20\leq {\cal C} \leq 20$ (the density was
progressively increased until the final roots did not change. The
final density we used was such that the $\mu^2$ values are varied
in steps of $0.2$ and $\cal C$ values are varied in steps of
$0.02$ in every iteration). In our opinion, this range is sufficiently large to probe the stability
of the model against tachyonic modes. Indeed, the first KK modes are expected to have a mass
of the order of the inverse compactification radius, corresponding to $\vert \mu^2 \vert$ of order one.
Also the parameter ${\cal C}$ is naturally expected to be of order one (since it comes from a Taylor expansion). Moreover, Newton's method can converge to solutions outside this range of starting values for $m^2$ and ${\cal C}$ (this is indeed what happened in some cases). In all the cases studied, the two
distances $d_1$ and $d_2$ strongly increased at negative $m^2$
(they were typically several orders of magnitudes greater than for
positive $m^2$), indicating that no tachyons are present in the
model; to have a further check, we started Newton's method also
from some of the guessed values with negative $m^2$ (despite the
corresponding distances $d_1$ and $d_2$ were always very high); the
method never converged to any tachyonic mode.

\begin{figure}
\centering
\includegraphics[width=3in]{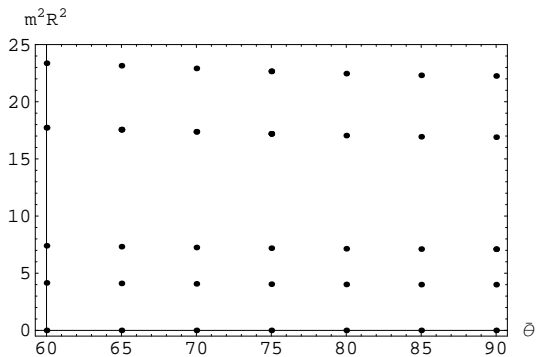}
\caption{Smallest masses for the vector modes, for various brane positions and for the specific value $\beta = 0.9 \,$.} \label{fig2}
\end{figure}
In figure \ref{fig2} we show the obtained spectrum for several
brane positions and for the specific choice $\beta=0.9$. We note
the presence of one massless mode. As for the tensor case, we also
show the results for the strict codimension two case (${\bar
\theta} = 90^0$). We observe that, also for the vector sector, the
limit of shrinking the extended brane to a codimension two defect
is continuous (this can also be proven analytically from the study
of the linearized equations, in the same way as we did for the
tensor modes). Moreover, also for the vector sector, we observed a
weak dependence (not shown here) of the spectrum on the value of
the deficit angle $\beta$ (the reason is the same as for the
tensor sector).

\subsection{Scalar Modes}

There are four scalar gauge invariant combinations, satisfying the set of linearized equations
derived in~\cite{PST}. Two bulk constraint equations (containing at most first order derivatives) can
be used to express ${\hat C}$ and ${\hat a}_\phi$ in terms of the other two modes ${\hat \Phi}$ and $\Psi \,$. This leaves us with the two bulk equations
\begin{eqnarray}
&& \frac{d^2 {\hat \Phi}}{d^2 \theta} + \tan \theta \, \left( 4 \frac{d \Psi}{d \theta} - 3 \frac{d {\hat \Phi}}{d \theta} \right) + \left( \frac{3\mu^2}{2} - 2\right) {\hat \Phi} - \frac{\mu^2}{2} \Psi=0
\nonumber\\
&& \frac{d^2 \Psi}{d \theta^2} + \tan \theta \, \frac{d \Psi}{d \theta} + \frac{\mu^2}{2} \Psi+\frac{\mu^2}{2} {\hat \Phi} = 0
\label{bulksca}
\end{eqnarray}
where $\mu$ is related to the physical mass as in~(\ref{eqh}).

The parity assignment of the background imposes that both modes are even. Moreover, regularity
at the poles requires
\begin{equation}
\hat{a}_{\hat\phi} \,_{| \pm \pi/2} = \frac{d \hat{a}_{\hat\phi}}{d \theta} \,_{| \pm \pi/2} =
\frac{d \Psi}{d \theta} \,_{| \pm \pi/2} =
\frac{d {\hat \Phi}}{d \theta} \,_{| \pm \pi/2} = 0
\end{equation}
Once we insert these conditions in one of the constraint bulk equations (which is legitimate, since
the involved quantities are continuous as we approach the poles), we find
\begin{equation}
\left( {\hat \Phi} + \Psi \right) \,_{| \pm \pi/2}=0
\end{equation}

The junction conditions at the brane location were also expressed in~\cite{PST} in terms of all $4$ gauge invariant scalar combinations, plus the the quantity $E' \,$. This quantity can be identified with the (scalar) perturbation of the brane position. In an arbitrary gauge, the brane is at the perturbed position $\theta = {\bar \theta} + \zeta \left( x^\mu \right) \,$. This quantity changes when we performed a change of coordinates involving the bulk coordinate $l \,$. The combination which does not change under such change of coordinate is ${\hat \zeta} = \zeta - E' \,$, which can be then interpreted as the gauge invariant perturbation of the brane position. Not surprisingly, this is the quantity which enters in the junction conditions, when they are written in terms of the gauge invariant perturbations (\ref{hatsca}). We further restricted the gauge freedom by choosing a system of coordinates where the brane remains at the background position, that is $\zeta = 0 \,$. In this case, ${\hat \zeta} = - E' \,$, which is the quantity entering in the junction conditions.

We can combine the junctions conditions given in~\cite{PST} to eliminate $E' \,$. This requires the use of the bulk equations (which is however legitimate, since the junction conditions relate bulk quantities at the two sides of the brane). After some algebra, we find
\begin{eqnarray}
&& [\Psi]_{J}=0 \nonumber\\
&& [(1+\sin^2\bar\theta)\hat\Phi+\sin\bar\theta\cos\bar\theta(\Psi'-\hat\Phi')]_{J}=0
\nonumber\\
&& \left[\theta'\left(4\Psi'-\mu^2\cos\bar\theta\sin\bar\theta(\Psi-\hat\Phi)\right)\right]_{J}=0
\nonumber\\
&&
\left[\left(-5\Psi'-\tan\bar\theta\hat\Phi+\hat\Phi'\right)\theta'-\frac{1+1/\beta}{\cos\bar\theta\sin\bar\theta}\hat\Phi\right]_{J}=0 \nonumber\\
\label{juncsca}
\end{eqnarray}
where $\left[ f \right]_J \equiv f_{\rm out} - f_{\rm in}$ denotes the difference of the quantity $f$
between the two sides of the brane.

Also for the scalar sector, the bulk equations~(\ref{bulksca})
must be solved numerically. We therefore perform a numerical
analysis analogous to the one done for the vector modes. We first
solve the bulk equations analytically in a neighborhood of the
south pole. Fixing the overall normalization by setting $\Psi =1$
at the south pole~\footnote{This choice does not include the possibility of $\Psi = 0$ at the pole (such modes could in principle exist, since they could have a nonvanishing second derivative at the pole). We performed a separate numerical investigation for this case, which however did not show the  existence of any such mode.}, and taking into account the regularity
conditions mentioned above, we find
\begin{eqnarray}
\Psi \left( - \pi/2 + \epsilon \right) &=& 1 + {\cal C} \, \epsilon^2 + {\rm O} \left( \epsilon^4 \right) \nonumber\\
{\hat \Phi} \left( - \pi / 2 + \epsilon \right) &=& - 1 + \frac{\mu^2 + 4 \, {\cal C} - 1}{4} \epsilon^2 + {\rm O} \left( \epsilon^4 \right)
\label{scasmalleps}
\end{eqnarray}

\begin{figure}
\centering
\includegraphics[width=3in]{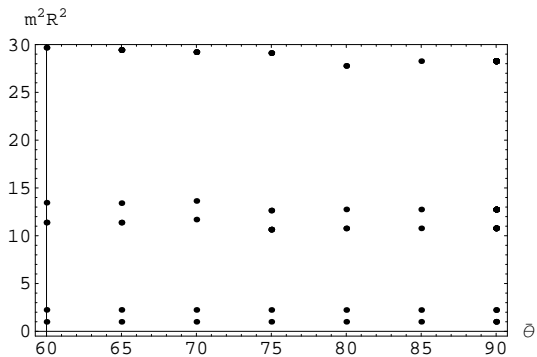}
\caption{Smallest masses for the scalar modes, for various brane positions and for the specific choice $\beta = 0.9 \,$.} \label{fig3}
\end{figure}

Also in this case, the mode is uniquely determined by the two
parameters $m^2$ and ${\cal C}$. The numerical investigation then
proceeds in the same way as for vectors. Fig. \ref{fig3} shows the
lightest masses in the spectrum, for the specific choice of
$\beta=0.9$ and for different brane positions (the small ``oscillatory'' behavior of the 
eigenmasses visible in the figure is probably due to numerical errors, and it 
gives a measure of the precision of the computation). As for the the
other two sectors, the computation does not show any tachyonic
modes (the modes exhibit a very bad parity for all negative values
of $m^2$ we have attempted). However, there are two interesting
differences between the scalar and the other two sectors.

The first difference is the absence of scalar zero modes (as can
be also verified by solving the equations analytically, which is
possible for vanishing mass). This indicates that all the moduli
of the model have been lifted (by the fluxes and tensions in the
system), and that the compactification is stable (this is the case also for a 
codimension $2$ brane at the pole~\cite{graesser}). The present
stability analysis is done in absence of (matter or gauge) fields
localized on the brane. Ref.~\cite{PST} studied the gravitational
interaction between brane sources; it was found that two zero
modes are then excited, and contribute to reproduce Einstein $4$d
gravity at large distances. A similar situation is also
encountered in $5$d models, for instance the Randall-Sundrum model
with a single brane~\cite{rs2}. The background solution
of~\cite{rs2} has no scalar perturbations; however, when (matter)
fields are localized on the brane, a scalar zero mode - often
denoted as brane bending~\cite{Garriga} - is excited, and gives a relevant
contribution to the gravitational interactions between the brane
fields.

The second peculiarity of the scalar sector is that the limit
${\bar \theta} \rightarrow \pi /2$ is discontinuous. To see this, we solve
the linearized equations when the brane is close to the pole,
at the position $\theta = - \pi/2 + \epsilon \,$. 
The analytical solution (\ref{scasmalleps}) then accurately
describes the modes in the in bulk immediately before the brane. We
can then expand the junction conditions (\ref{juncsca}) for small
$\epsilon$, and obtain the values of the modes in the out part of
the bulk immediately after the brane. They are
\begin{eqnarray}
\Psi_{\rm o} &=& 1 + {\rm O} \left( \epsilon^2 \right) \nonumber\\
\frac{d \Psi_{\rm o}}{d \theta} &=& \frac{8 \, {\cal C} + 3 \left( \beta - 1 \right) \beta m^2}{4 \beta} \, \epsilon
+ {\rm O} \left( \epsilon^3 \right) \nonumber\\
{\hat \Phi}_{\rm o} &=& - 2 + \beta + {\rm O} \left( \epsilon^2 \right) \nonumber\\
\frac{d {\hat \Phi}_{\rm o}}{d \theta} &=& \frac{2 \left( 1 - \beta \right)}{\epsilon} + {\rm O} \left( \epsilon \right)
\label{outsmall}
\end{eqnarray}
The limit $\epsilon \rightarrow 0$ (i.e. ${\bar \theta} \rightarrow \pi /2$) would be continuous if, these values converged to those which must be imposed for a codimension two brane at the pole. The latter values are $\Psi + {\hat \Phi} = d \Psi / d \theta = d {\hat \Phi} / d \theta = 0 \,$, which clearly shows that the limit is not continuous (the only exception is the trivial case of a vanishing deficit angle, $\beta = 1$, when both cases collapse to a spherical compactification with an empty brane at the pole).

This discontinuity, however, does not lead to any appreciable discontinuity on the lowest eigenmasses, as can be observed from fig. \ref{fig3} (the values for ${\bar \theta} = \pi /2$ refers to the codimension two brane). For small $\epsilon \,$, the eigenfunctions, although starting from a different value on the outside bulk, quickly approach the ones of the codimension two case, leading to nearly identical eigenmasses (within the accuracy of the numerical computation).

We actually observe from the last of~(\ref{outsmall}) that the limit ${\bar \theta} \rightarrow \pi /2$ actually leads to a divergent derivative of ${\hat \Phi}$ on the outside bulk. This results in divergent terms in the linearized Einstein tensor. An analogous result is also encountered for the scalar modes excited by brane fields. It was found in~\cite{PST} that one scalar mode diverges when the codimension one brane is shrunk to the pole. This singular limit in the scalar sector is what precludes the localization of matter fields on a strict codimension two brane.

\section{Conclusions}

We studied a brane-world model in a six dimensional space-time, in which two of the dimensions are compactified by a flux. The model is characterized by a codimension one brane, with one dimension 
extending inside the compact space. This construction avoids the singularities which plague codimension two and higher defects, where only tension can be localized. Indeed, it was shown in~\cite{PST} that fields with arbitrary tension can be localized on this defect, and that their gravitational interaction is described by Einstein $4$d gravity at large distance. 

We do not regard this construction as a regularized codimension two brane, in the sense that one does not recover a well-behaved solution as the extended defect is shrunk to a zero size in the bulk. More accurately, the linearized computation of gravity in this model breaks down (in the scalar sector) as the size of the brane decreases. While the only safe claim that one can make is that gravity becomes strong in this limit, it is hard to expect a regular nonperturbative limit, in the light of the fact that the strict codimension two case is itself badly singular. The positive aspect of this statement is that this construction leads to distinct and potentially testable predictions for short-range gravitational and electroweak interactions (corrections to standard gravity would show up at distances comparable to the compactification radius, while electroweak effects would appear at energies close to the inverse size of the brane in the internal space). We expect that other inequivalent regular constructions are possible, but that they would lead to different short range observables.

To make clear predictions, one needs to obtain the massive spectrum of perturbations of the model. While only the massless modes were studied in~\cite{PST}, we performed this computation in the second part of this note. Even more importantly, the computation is mandatory to verify that no tachyonic mode is present, so that the construction is stable. As also argued in~\cite{burgess}, the study of axially symmetric perturbations should be enough for this check (since the background is itself axially symmetric, modes with a nontrivial angular dependence are expected to have a higher mass). For this reason, we focused our investigation on these modes.

Due to technical difficulties that we have discussed in the paper, we were not able to decouple the system of bulk equations for the vector and scalar modes, so that we had to resort to a numerical investigation. We did so with a shooting method (slightly modified, to cope with the coordinate singularities at the poles). Clearly, numerical methods can only guarantee the stability within the range probed. However, we conducted a rather extensive search. While Kaluza-Klein masses are naturally expected to be of the order of the inverse compactification radius, we densely investigated a parameter space in the interval~\footnote{We actually employed a Newton method which could converge to solutions also outside this interval, if present.} $- 100 / R^2 \leq m^2 \leq 100 / R^2 \,$, and for several bulk parameters (brane position and deficit angle). In no case we found evidence for tachyonic solutions. Since this is a relatively easy numerical problem (the bulk is regular, and there are no strong hierarchies  present), we believe that the present analysis ensures the stability of the construction.

While the above considerations are valid for a Minkowski external space, in the first part of this analysis we studied the dS solutions of this model. We found that the space of vacua is characterized by discrete points labeled by two integer numbers, related to the quantized values of the flux in the bulk ($N$) and of the current of a brane field ($n$), which, in turns, controls the position of the brane in the internal space. The Minkowksi compactification requires $N = - 2n$. If this is the case, we can actually have different Minkowski compactifications, provided the cosmological constant on the brane ($\lambda$) and on the two sides of the bulk ($\Lambda_{i,0}$) satisfy
\begin{equation}
\sqrt{\lambda} = \frac{\sqrt{\Lambda_i} - \sqrt{\Lambda_0}}{q} \;\;,\;\; \Lambda_i = \frac{2 e^2 M^8}{N^2}
\label{mink}
\end{equation}
As long as these relations are satisfied, a change in the brane and bulk tensions leads to a different
internal space, but only to Minkowski external geometry. Since the ratio of the two integers $N$ and $n$ cannot be varied continuously, it is tempting to ask whether this can be of some relevance for the cosmological constant problem.

Discrete vacua typically arise in presence of fluxes, and several studies have already attempted  to use this as a solution of the cosmological constant problem.  The original mechanism of~\cite{BT} is realized with a $4-$form in four dimensions. This form can acquire only quantized values in units of a charge $q$, and its energy density behaves as a cosmological constant. The value of the form can change through membrane nucleation; this is however a very slow process, and it could be possible that the present universe is trapped in a metastable state, where the vacuum energy and of the $4-$form add up to the observed value of the cosmological constant $\Lambda_{\rm tot}$. This mechanism allows for several possible values of $\Lambda_{\rm tot}$, and, provided these values are densely packed together, one may hope to reproduce the observed expansion rate for some value of the quantized flux 
(even if this is not the case initially, one should simply wait until the flux tunnels to a value compatible with observations). This, however, requires very small values of $q$, which - besides being unnatural - do not lead to the known cosmology (they would lead to a reheating temperature much smaller than the one needed for primordial Nucleosynthesis, see~\cite{BP} for a discussion). String theory can offer a big improvement in this respect. There, one requires an internal space which can be stabilized by fluxes~\cite{KKLT}, and which can have a complex topology, with several non--contractible cycles. There are various ways of wrapping fluxes around such internal space, leading to several quantized fluxes, and to a multi--dimensional set of discrete vacua. Due to the high dimensionality, it is natural to expect vacua with a value of $\Lambda_{\rm tot}$ compatible with the observed one, even if the distance between different vacua (namely, the value of the charge $q$) is large~\cite{BP}. This is one of the possible realizations of the landscape of string theory~\cite{sus}.

In the present context, inserting the measured value of $H$ in eq.~(\ref{dsratio}), we find
\begin{equation}
1 + \frac{2 n}{N} \sim 10^{-60} \left( \frac{R}{0.1 \, {\rm mm}} \right)^2
\label{final}
\end{equation}
In the mechanism of~\cite{BT}, the present value of $H$ is achieved at the price of an unnaturally small
charge, and very large flux. Our realization does not put a significant constraint of the charge. However, the value~(\ref{final}) requires a very tiny ``mismatch'' of the relation $2 n / N = -1$, which can be achieved only when the two winding numbers are themselves ${\rm O} \left( 10^{60} \right)$. Although 
the corresponding request (very large flux) is usually not listed as a drawback of~\cite{BT}, it is hard to regard such high windings as natural. 

We can think of some possible improvement. The necessity of large windings  is probably due to the extreme simplicity of the model. As we mentioned, the problem of localizing fields in general relativity is present for any defect of codimension higher than one, and not just for codimension two. Assuming that such a construction must be done also for more realistic (and richer) models, we can expect that the presence of more fluxes can allow for a solution without too large winding, in a similar way as the string realization improves over the one of~\cite{BP}. Alternatively, 
we may be satisfied in providing at least a partial solution to the cosmological constant problem. Rather than requiring large windings, it is probably more natural to assume that we are locked in one of the Minkowski vacua, with $N = - 2n$. This relation (which by itself does not appear unnatural) could possibly explain why the ``big'' cosmological constant vanishes. The coincidence problem could instead be solved by some additional field, which would then play the role of quintessence. This would still be an improvement with respect to  the usual models of quintessence, where the absence of a ``big'' cosmological constant is typically left unexplained. 

To improve over these considerations requires a better understanding of the background solutions of the model. For instance, it may be possible to have $N = - 2 n$, even if the cosmological constants do not satisfy the relations~(\ref{mink}). This may be compatible with a more general solution than the restricted ansatz (Minkowski or dS external geometry, times a static internal space) assumed here. Moreover, in order to study quintessence in this context, one needs to include sources with a different equation of state than vacuum. It is usually hard, if not impossible, to obtain analytical solutions with a time evolving internal space. However, such questions can be possibly addressed analytically at low energies, or numerically along the lines of~\cite{branecode}.

\vskip 0.5in
\vbox{
\noindent{ {\bf Acknowledgments} } \\ \\
\noindent
We thank Tony Gherghetta, Nemanja Kaloper, Lorenzo Sorbo, and Gianmasimo Tasinato for very useful discussions. This work was partially supported by DOE grant DE--FG02--94ER--40823, and by a grant from the Office of the Dean of the Graduate School of the University of Minnesota.  
}

\end{document}